\documentclass{article}
\usepackage{fullpage,url}

\title{\bf $k$-Anonymously Private Search over Encrypted Data}
\author{Shiyu Ji, Kun Wan\\\{shiyu, kun\}@cs.ucsb.edu}
\date{}

\usepackage{graphicx}
\usepackage[linesnumbered,ruled]{algorithm2e}

\begin{document}
\newcommand{\Enc}{\textsf{Enc}}
\newcommand{\Dec}{\textsf{Dec}}

\maketitle

\begin{abstract}
    In this paper we compare the performance of various homomorphic encryption methods on a private search scheme that can achieve $k$-anonymity privacy. To make our benchmarking fair, we use open sourced cryptographic libraries which are written by experts and well scrutinized. We find that Goldwasser-Micali encryption achieves good enough performance for practical use, whereas fully homomorphic encryptions are much slower than partial ones like Goldwasser-Micali and Paillier.
\end{abstract}

\section{Introduction}
How to securely and privately search over encrypted dataset has been popular in research community for the past decade \cite{KO97,KO00,Lip05,Lip09,cash2013highly, cash2014dynamic,ABF16}.
A private search scheme should allow the client to have access to one of the posting lists in the inverted index, while the server hosting the index cannot tell which posting is accessed by the client. Also for privacy concern, the client should not know any other posting except the legally requested one. 
To address this problem, one possible way is to leverage Homomorphic Encryption \cite{GM82,GM84,Pai99a,Pai99b,gentry2013homomorphic,BGV12,BGV14}. 
Recently there is some debate over the question whether such a homomorphic encryption based private search scheme can be practically used \cite{SC07,OG11,ABF16}. We note that it is important which encryption method is used, since the involved operations may be very different, e.g., there are many modular exponential operations in Paillier \cite{Pai99a,Pai99b}, while Goldwasser-Micali only requires quadratic exponentiation; some lattice based HEs (e.g. \cite{BV11,BGV12}) require noise management techniques such as bootstrapping \cite{Gen09}, re-linearization \cite{BV11} and approximate eigenvector \cite{GSW13}, which need more time. It will be interesting to see which HE method outperforms others in performance.

The main objective of this paper is to compare the performance on private search given by existing homomorphic encryption methods. The main challenge is that to write a cryptographic library is time-consuming and error-prone. Thus to do a fair comparison, we need well written libraries which are publicly available and have been scrutinized by experts. Fortunately there are many published libraries available for Homomorphic Encryption \cite{gitLibgm,gitIvan,gitHS,gitDM}, and many of them are written by the original paper authors \cite{gitHS,gitDM}. We will do a benchmark for each library and compare the performance results. 

\section{Preliminaries and Related Works}

\subsection{Homomorphic Encryption}
Homomorphic encryption (HE) is an encryption system that allows some computations over ciphertexts without knowing the plaintexts or secret keys. A homomorphic encryption that allows the computation of any circuit (including additions and multiplications) over ciphertexts is called Fully Homomorphic Encryption (FHE). If the computation scope is restricted (e.g., only addition or multiplication by some constant), it is called Partially Homomorphic Encryption (PHE).
\begin{itemize}
    \item \emph{Partially Homomorphic Encryption}: There is a line of research works that can support HE with any times of additions: ElGamal \cite{Elg85}, Goldwasser-Micali \cite{GM82,GM84}, Benaloh \cite{Bena94}, Paillier \cite{Pai99a,Pai99b}, etc. A PHE $\Enc$ supporting addition can compute $\Enc(m_1 + m_2)$ only given the ciphertexts $\Enc(m_1)$ and $\Enc(m_2)$. Furthermore, Goldwasser-Micali can support the muliplication of two bits, i.e., to compute $\Enc(b_1b_2)$ only given $\Enc(b_1)$ and $\Enc(b_2)$ where $b_1, b_2 \in \{0, 1\}$. Paillier can support multiplication by constant, i.e., to compute $\Enc(m_1m_2)$ only given $\Enc(m_1)$ and $m_2$ where $m_1$, $m_2$ are plaintexts. However, PHE \emph{cannot} support addition and multiplication over ciphertext simultaneously.
    \item \emph{Fully Homomorphic Encryption}: Since the first time to be proposed by Craig Gentry \cite{Gen09}, there is a long line of research works on FHE \cite{VGH10,SV10,SS10,Gen10,GH11a,GH11b,BV11,BGV12,BLL13,GSW13,BGV14,BV14}. So far most FHE schemes are built on difficult lattice problems \cite{Reg09}. By FHE, one may arbitrarily compute over ciphertexts, e.g., add and multiply for many times. The major challenge in the design of FHE is noise management, i.e., the noise in ciphertext grows rapidly when multiplying \cite{BV11,BV14}. Thus some time-consuming techniques like squashing \cite{Gen09,Gen10,GH11a,GH11b,LM15}, bootstrapping \cite{Gen09}, re-linearization \cite{BV11, BV14}, approximate eigenvector \cite{GSW13} are needed. The difficulty of noise management makes it very challenging to improve the performance of FHE in practice, and there is a long way to go to make an efficient FHE implementation by industry standard \cite{NLV11}.
\end{itemize}

\subsection{Private Information Retrieval and Private Search}
A private information retrieval (PIR) scheme allows the client to obtain a posting in the inverted index from the server, while the server cannot learn which posting is requested by the client. Clearly the most trivial way is to give all the postings to the client. This takes too much bandwidth, and the client learns everything: if it is a dataset of medical records, the client learns the private information from anyone else. 
Hence a private search scheme further requires that the client cannot learn any other information except the requested posting.

There is a line of research on PIR \cite{KO97,CMS99,KO00,Lip05,Lip09}. These works tried to reduce the computation time and communication overhead of PIR by assuming some computational intractability, e.g., quadratic residuosity problem \cite{GM82,GM84}, learning with errors \cite{Reg09}, etc. So far the best result given by \cite{Lip09} is that the computation time is $O(n/\log n)$ and the communication overhead is $O(\log^2 n)$ where $n$ is the size of dataset and the client requests one bit in the dataset.

Note that a PIR scheme that has absolutely no leakage is widely considered to be inefficient in practice (e.g., the response time can be 10 to 1000 seconds) \cite{SC07,OG11}. However, if we allow some acceptable leakage, can we obtain an efficient PIR scheme that can be used in the real world? In this paper we investigate this possibility.

\subsection{$k$-Anonymity Privacy}
As a standard notion of privacy, together with its variants \cite{MKG07,LLV07}, $k$-anonymity has been researched for many years \cite{Swe02a,Swe02b,LD05,LD06,MKG07,LLV07}. A key-value based dataset has $k$-anonymity if the value for each key cannot be distinguished from at least $k-1$ other values whose keys also appear in the dataset. For a $k$-anonymous search scheme, for each request from the client, there always exist $k$ entries in the inverted index such that the server cannot tell which one of these $k$ entries is requested by the client. There is absolutely no leakage if $k$ equals the number of entries in the inverted index.

\section{A Private Search Scheme Achieving $k$-Anonymity}
In this paper we investigate the performance of the private search scheme given as Algorithm \ref{alg:ps}. The HE based idea of this scheme is similar to \cite{KO97, KO00, Lip05, Lip09, ABF16}.

\newcommand{\pk}{\textsf{pk}}
\newcommand{\sk}{\textsf{sk}}
\begin{algorithm}
\SetAlgoLined
 \KwData{$D$: the inverted index containing $k$ posting lists: $D_1$, $\cdots$, $D_k$\; $i$: the index of the posting that the client requests; $p$: the length of posting list, i.e., $p = |D_j|$ for any index $j$\; $\Enc$: a PHE supporting addition and multiplication by constant with public key $\pk$ and secrete key $\sk$.}
 \KwResult{The client learns $D_i$ and nothing more, while the server cannot tell $i$.}
 For each index $j \in [1, k]$, the client computes $c_j = \Enc_{\pk}(b_j)$ where $b_j$ equals to 1 if $i=j$ and 0 otherwise\;
 The client sends $C = \{c_1, \cdots, c_k\}$ to the server\;
 For each offset $s \in [1, p]$, the server computes $e_s = \sum_{j=1}^k c_j \cdot D_j[s]$, where $D_j[s]$ denotes the $s$-th bit in the posting $D_j$\;
 The server sends $E = \{e_1, \cdots, e_p\}$ to the client\;
 The client decrypts each ciphertext $e_s$ in $E$ by using $\sk$ to recover the posting list $D_i$\;
 \caption{A $k$-anonymous private search scheme.}
 \label{alg:ps}
\end{algorithm}

Clearly the computation time of the scheme as above is $O(kp)$ and the communication overhead is $O(k+p)$. The correctness can be verified as follows.
$$\Dec_{\sk}(e_s) = \Dec_{\sk}(\sum_{j=1}^k \Enc_{\pk}(b_j) \cdot D_j[s])
= \Dec_{\sk}(\Enc_{\pk}(\sum_{j=1}^k b_j \cdot D_j[s])) = \Dec_{\sk}(\Enc_{\pk}(D_i[s])) = D_i[s].$$
For practical use, very often we do not need too large $k$ (e.g., at most 100). However, the number of entries in the inverted index can be much larger. One possible solution is to divide the inverted index into several blocks with size $k$, and the client lets the server know which block will be requested. A leakage immediately follows since now the server knows which block is frequently requested, and thus the search pattern is not completely hidden. But if we only require $k$-anonymous, then the postings within one block are $k$-anonymous since the server cannot tell which one of them is requested.

\section{Evaluation Results}
In this section we summarize the evaluation results. By running similarly implemented testing code, we run the homomorphic encryption based private search schemes using the libraries as follows:
\begin{itemize}
    \item Praveen Kumar's C library \emph{libgm} \cite{gitLibgm} implementing the Goldwasser-Micali cryptosystem.
    \item Gustavo Brunoro's Python code \cite{gitBru} implementing the Goldwasser-Micali cryptosystem. We denote this code as PyGM.
    \item Mike Ivanov's Python library \cite{gitIvan} implementing the Paillier cryptosystem.
    \item Our Python code based on \cite{gitBru} implementing the Cachin-Micali scheme \cite{CMS99}.\footnote{Cachin-Micali scheme is a PIR scheme, \emph{not} a HE system.}
    \item Shai Halevi and Victor Shoup's HElib \cite{gitHS} (written in C++) implementing the Brakerski-Gentry-Vaikuntanathan (BGV) scheme \cite{BGV12,BGV14}.
    \item Leo Ducas and Daniele Micciancio's FHEW \cite{gitDM} (written in C++) implementing their own FHE scheme \cite{LM15}.
\end{itemize}
For the schemes which need large primes, we generate 2048-bit primes, which are considered to be secure so far \cite{Kal03}. For the lattice or LWE based FHE schemes, we use the minimum security configuration given by the authors. We choose the posting length to be 720 bits. We consider different $k$'s in $k$-anonymity when evaluating.

We run the experiments on Linux Ubuntu 16.04 servers with 8 cores of 2.4 GHz AMD FX8320, 16GB memory. Since some libraries do not provide enough interfaces (e.g., except for encryption and decryption, FHEW only provides ciphertext evaluation for NAND to use), we cannot guarantee that all of our implementation are optimal.

\subsection{Key Generation Time}

\begin{table}
\centering
\begin{tabular}{|c|c|c|c|c|c|}
    \hline
    libgm & PyGM & Paillier & Cachin-Micali & HElib & FHEW \\
    \hline
    0.84 & 0.015 & 0.024 & 0.40 & 6.13 & 14.8 \\
    \hline
\end{tabular}
\caption{Key generation time (sec) for different $k$'s.}
\label{tab:kgen}
\end{table}

Table \ref{tab:kgen} gives the key generation time results. Goldwasser-Micali and Paillier are the best. Note that libgm uses GNU MP Bignum library, while PyGM uses Miller-Rabin to test primality. GNU MP generates large primes that are safer to use, but takes more time. 

\subsection{Query Encryption Time}

\begin{table}
\centering
\begin{tabular}{|c||c|c|c|c|c|c|}
    \hline
    $k$ & libgm & PyGM & Paillier & Cachin-Micali & HElib & FHEW \\
    \hline
    10  & 0.02 & 0.0002 & 0.93 & 0.40 & 1.03 & 6.8e-5 \\
    20  & 0.03 & 0.0004 & 1.86 & 0.57 & 2.04 & 1.3e-4 \\
    50  & 0.07 & 0.0007 & 4.59 & 1.59 & 5.18 & 3.3e-4 \\
    100 & 0.15 & 0.0012 & 19.90 & 3.02 & 15.76 & 6.7e-4 \\
    \hline
\end{tabular}
\caption{Query encryption time (sec) for different $k$'s.}
\label{tab:enc}
\end{table}

Table \ref{tab:enc} gives the query encryption time results. Goldwasser-Micali and FHEW achieve the best performance.

\subsection{Query Execution Time}
The query execution includes the time that the server uses to generate the ciphertext to return based on the query, and the decryption time on the client side.

\begin{table}
\centering
\begin{tabular}{|c||c|c|c|c|c|c|}
    \hline
    $k$ & libgm & PyGM & Paillier & Cachin-Micali & HElib & FHEW \\
    \hline
    10  & 0.172 & 0.113 & 0.307 & 0.43 & 1.61 & 16k \\
    20  & 0.175 & 0.130 & 0.316 & 0.81 & 3.08 & 33k \\
    50  & 0.185 & 0.164 & 0.361 & 1.87 & 7.70 & 82k \\
    100 & 0.214 & 0.232 & 0.440 & 6.09 & 20.93 & 164k \\
    \hline
\end{tabular}
\caption{Query execution time (sec) for different $k$'s.}
\label{tab:exe}
\end{table}

Table \ref{tab:exe} gives the query execution time results. Goldwasser-Micali scheme achieves the best performance. Note that FHE schemes (HElib and FHEW) take more time on the additions and multiplications over ciphertexts, since the noise in ciphertexts grows rapidly by multiplication and hence the scheme needs to control the noise.

\subsection{Communication Overhead}

\begin{table}
\centering
\begin{tabular}{|c||c|c|c|c|c|c|}
    \hline
    $k$ & libgm & PyGM & Paillier & Cachin-Micali & HElib & FHEW \\
    \hline
    10  & 11680 & 6440 & 6440 & 6528 & 1408 & 22k \\
    20  & 11840 & 6512 & 6512 & 6600 & 2688 & 42k \\
    50  & 12320 & 6776 & 6776 & 6864 & 6528 & 102k \\
    100 & 13120 & 7160 & 7160 & 7248 & 12928 & 202k \\
    \hline
\end{tabular}
\caption{Communication overhead (bytes) for different $k$'s.}
\label{tab:co}
\end{table}

Table \ref{tab:co} gives the communication overhead results. We evaluate communication overhead as the number of bytes communicated between the client and server during the private search protocol. Goldwasser-Micali, Cachin-Micali and Paillier achieve the best performance, and their overhead grows slowly as $k$ increases.

\section{Conclusion and Future Works}
Based on the benchmarking results, Goldwasser-Micali encryption achieves good enough performance for practical use, whereas fully homomorphic encryptions are much slower than partial ones like Goldwasser-Micali and Paillier. We suggest that it may be worthwhile to research on the possibility of incorporating Goldwasser-Micali method into private search scheme in the future works.

\bibliographystyle{plain}
\bibliography{references}
\end{document}